\magnification=\magstephalf
\font\t=cmcsc10 at 13 pt

\font\n=cmcsc10
\font\foot=cmr9

\centerline{\t On the Gravitational Field of a Sphere}
\smallskip
\centerline{\t of Incompressible Fluid}
\smallskip
\centerline{\t according to Einstein's Theory \footnote{\dag}
{\foot{Sitzungsberichte der K\"oniglich Preussischen Akademie
der Wissenschaften zu Berlin, Phys.-Math. Klasse 1916, 424-434.}}}\bigskip
\centerline{\n by K. Schwarzschild}\bigskip
\centerline{(Communicated February 24th, 1916 [see above p.
313].)}\par\vbox to 1.0 cm {}

\centerline{(Translation\footnote{\ddag}{\foot The valuable
advice of A. Loinger is gratefully acknowledged.}
by S. Antoci\footnote{$^*$}{\foot Dipartimento di Fisica
``A. Volta'', Universit\`a di Pavia, Via Bassi 6 - 27100 Pavia
(Italy).})}\bigskip

\S 1. As a further example of Einstein's theory of gravitation I
have calculated the gravitational field of a homogeneous sphere of
finite radius, which consists of incompressible fluid. The addition
``of incompressible fluid'' is necessary, since in the theory of
relativity gravitation depends not only on the quantity of matter,
but also on its energy, and {\it e. g.} a solid body in a given
state of tension would yield a gravitation different from a
fluid.\par
The computation is an immediate extension of my communication on
the gravitational field of a mass point (these Sitzungsberichte
1916, p. 189), that I shall quote as ``Mass point'' for short.\par
\S 2. Einstein's field equations of gravitation (these
Sitzungsber. 1915, p. 845) read in general:

$$\sum_\alpha{\partial \Gamma^\alpha_{\mu\nu}\over {\partial
x_\alpha}}
+\sum_{\alpha\beta}~\Gamma^\alpha_{\mu\beta
}\Gamma^\beta_{\nu\alpha}=G_{\mu\nu}.\eqno(1)$$
The quantities $G_{\mu\nu}$ vanish where no matter is present. In the
interior of an incompressible fluid they are determined in the
following way: the ``mixed energy tensor'' of an incompressible fluid
at rest is, according to Mr. Einstein (these Sitzungsber. 1914,
p. 1062, the $P$ present there vanishes due to the incompressibility):

$$T^1_1=T^2_2=T^3_3=-p,~~~T^4_4=\rho_0,
~~~(the~remaining~T^\nu_\mu=0).\eqno(2)$$
Here $p$ means the pressure, $\rho_0$ the constant density of the
fluid.\par
The ``covariant energy tensor'' will be:

$$T_{\mu\nu}=\sum_\sigma T^\sigma_\mu g_{\nu\sigma}.\eqno(3)$$
Furthermore:

$$T=\sum_\sigma T^\sigma_\sigma=\rho_0-3p\eqno(4)$$
and

$$\kappa=8\pi k^2,$$
where $k^2$ is Gauss' gravitational constant. Then according to Mr.
Einstein (these Berichte 1915, p. 845, Eq. 2a) the right-hand
sides of the field equations read:

$$G_{\mu\nu}=-\kappa(T_{\mu\nu}-{1\over 2}g_{\mu\nu}T).\eqno(5)$$
Since the fluid is in equilibrium, the conditions

$$\sum_\alpha{\partial T^\alpha_\sigma\over {\partial x_\alpha}}
+\sum_{\mu\nu}~\Gamma^\mu_{\sigma\nu }T^\nu_\mu=0\eqno(6)$$ must
be satisfied (ibidem Eq. 7a).\par
\S 3. Just as in ``Mass point'', also for the sphere
the general equations must be specialised to the case of rotation
symmetry around the origin. Like there, it is convenient to
introduce the polar coordinates of determinant $1$:

$$x_1={r^3\over 3},~~x_2=-\cos\vartheta,~~x_3=\phi,~~x_4=t.\eqno(7)$$
Then the line element, like there, must have the form:

$$ds^2=f_4dx_4^2-f_1dx_1^2-f_2{dx_2^2\over {1-x_2^2}}
-f_2dx_3^2(1-x_2^2),\eqno(8)$$
hence one has:

$$g_{11}=-f_1,~~g_{22}=-{f_2\over {1-x_2^2}},~~
g_{33}=-f_2(1-x_2^2),~~g_{44}=f_4$$
$$(the~remaining~g_{\mu\nu}=0).$$
Moreover the $f$ are functions only of $x_1$.\par
The solutions (10), (11), (12) reported in that paper hold also
for the space outside the sphere:

$$f_4=1-\alpha(3x_1+\rho)^{-1/3},~~f_2=(3x_1+\rho)^{2/3},~~
f_1f_2^2f_4=1,\eqno(9)$$
where $\alpha$ and $\rho$ are for now two arbitrary constants,
that must be determined afterwards by the mass and by the radius
of our sphere.\par
It remains the task to establish the field equations
for the interior of the sphere by means of the expression (8)
of the line element, and to solve them. For the right-hand
sides one obtains in sequence:

$$T_{11}=T^1_1g_{11}=-pf_1,~~T_{22}=T^2_2g_{22}=-{pf_2\over{1-x^2_2}},$$
$$T_{33}=T^3_3g_{33}=-pf_2(1-x^2_2),~~T_{44}=T^4_4g_{44}=\rho_0f_4.$$
$$G_{11}={\kappa f_1\over 2}(p-\rho_0),~~
G_{22}={\kappa f_2\over 2}{1\over {1-x^2_2}}(p-\rho_0),$$
$$G_{33}={\kappa f_2\over 2}(1-x^2_2)(p-\rho_0),~~
G_{44}=-{\kappa f_4\over 2}(\rho_0+3p).$$
The expressions of the components $\Gamma^\alpha_{\mu\nu}$ of the
gravitational field in terms of the functions $f$ and the left-hand
sides of the field equations can be taken without change from
``Mass point'' (\S 4). If one again restricts himself to the
equator ($x_2=0$), one gets the following overall system of
equations:\par
\noindent First the three field equations:
$$-{1\over 2}{\partial\over{\partial x_1}}\bigg({1\over f_1}
{\partial f_1\over{\partial x_1}}\bigg)+{1\over 4}{1\over
f_1^2}\bigg({\partial f_1\over{\partial x_1}}\bigg)^2+{1\over 2}
{1\over f_2^2}\bigg({\partial f_2\over{\partial x_1}}\bigg)^2
+{1\over 4}{1\over f_4^2}\bigg({\partial f_4\over{\partial x_1}}\bigg)^2
=-{\kappa\over 2}f_1(\rho_0-p),\eqno(a)$$
$$+{1\over 2}{\partial\over{\partial x_1}}\bigg({1\over f_1} {\partial
f_2\over{\partial x_1}}\bigg)-1-{1\over 2}{1\over{f_1f_2}}\bigg( {\partial
f_2\over{\partial x_1}}\bigg)^2
=-{\kappa\over 2}f_2(\rho_0-p),\eqno(b)$$
$$-{1\over 2}{\partial\over{\partial x_1}}\bigg({1\over f_1}
{\partial f_4\over{\partial x_1}}\bigg)
+{1\over 2}{1\over{f_1f_4}}\bigg( {\partial f_4\over{\partial
x_1}}\bigg)^2=-{\kappa\over 2}f_4(\rho_0+3p).\eqno(c)$$
In addition comes the equation for the determinant:

$$f_1f^2_2f_4=1.\eqno(d)$$
The equilibrium conditions (6) yield the single equation:

$$-{\partial p\over \partial x_1}=-{p\over 2}
\bigg[{1\over f_1}{\partial f_1\over \partial x_1}
+{2\over f_2}{\partial f_2\over \partial x_1}\bigg]
+{\rho_0\over 2}{1\over f_4}{\partial f_4\over x_1}.\eqno(e)$$
From the general considerations of Mr. Einstein it turns out that
the present 5 equations with the 4 unknown functions $f_1$, $f_2$,
$f_4$, $p$ are mutually compatible.\par
We have to determine a solution of these 5 equations that is free
from singularities in the interior of the sphere. At the surface
of the sphere it must be $p=0$, and there the functions $f$ together
with their first derivatives must reach with continuity the
values (9) that hold outside the sphere.\par
For simplicity the index $1$ of $x_1$ will be henceforth
omitted.\par
\S 4. By means of the equation for the determinant the equilibrium
condition (e) becomes:

$$-{\partial p\over\partial x}={{\rho_0+p}\over 2}{1\over f_4}
{\partial f_4\over\partial x}.$$
This can be immediately integrated and gives:

$$(\rho_0+p)\sqrt{f_4}=const.=\gamma.\eqno(10)$$
Through multiplication by the factors $-2$, $+2f_1/f_2$, $-2f_1/f_4$
the field equations (a), (b), (c) transform into:

$${\partial\over{\partial x}}\bigg({1\over f_1}
{\partial f_1\over{\partial x}}\bigg)=
{1\over{2f^2_1}}\bigg({\partial f_1\over{\partial x}}\bigg)^2
+{1\over{f^2_2}}
\bigg({\partial f_2\over{\partial x}}\bigg)^2
+{1\over{2f^2_4}}\bigg({\partial f_4\over{\partial x}}\bigg)^2
+\kappa f_1(\rho_0-p),\eqno(a')$$
$${\partial\over{\partial x}}\bigg({1\over f_2} {\partial
f_2\over{\partial x}}\bigg)
=2{f_1\over f_2}
+{1\over{f_1f_2}}{\partial f_1\over{\partial x}}
{\partial f_2\over{\partial x}}
-\kappa f_1(\rho_0-p),\eqno(b')$$
$${\partial\over{\partial x}}\bigg({1\over f_4}
{\partial f_4\over{\partial x}}\bigg)
={1\over{f_1f_4}}{\partial f_1\over{\partial x}}
{\partial f_4\over{\partial x}}
+\kappa f_1(\rho_0+3p).\eqno(c')$$
If one builds the combinations $a'+2b'+c'$ and $a'+c'$, by
availing of the equation for the determinant one gets:

$$0=4{f_1\over f_2}
-{1\over{f^2_2}}\bigg({\partial f_2\over{\partial x}}\bigg)^2
-{2\over{f_2f_4}}{\partial f_2\over{\partial x}}
{\partial f_4\over{\partial x}}+4\kappa f_1p\eqno(11)$$
$$0=2{\partial\over{\partial x}}\bigg({1\over f_2}
{\partial f_2\over{\partial x}}\bigg)
+{3\over{f^2_2}}\bigg({\partial f_2\over{\partial x}}\bigg)^2
+2\kappa f_1(\rho_0+p).\eqno(12)$$
We will introduce here new variables, which recommend themselves
since, according to the results of ``Mass point'', they behave
in a very simple way outside the sphere. Therefore they must bring
also the parts of the present equations free from $\rho_0$ and $p$
to a simple form. One sets:

$$f_2=\eta^{2/3},~~f_4=\zeta\eta^{-1/3},~~
f_1={1\over {\zeta\eta}}.\eqno(13)$$
Then according to (9) one has outside the sphere:

$$\eta=3x+\rho,~~\zeta=\eta^{1/3}-\alpha,\eqno(14)$$
$${\partial\eta\over \partial x}=3,~~
{\partial\zeta\over \partial x}=\eta^{-2/3}.\eqno(15)$$
If one introduces these new variables and substitutes
$\gamma f^{-1/2}_4$ for $\rho_0+p$ according to (10), the
equations (11) and (12) become:

$${\partial\eta\over \partial x}{\partial\zeta\over \partial x}
=3\eta^{-2/3}+3\kappa\gamma\zeta^{-1/2}\eta^{1/6}
-3\kappa\rho_0,\eqno(16)$$
$$2\zeta{\partial^2\eta\over \partial x^2}
=-3\kappa\gamma\zeta^{-1/2}\eta^{1/6}.\eqno(17)$$
The addition of these two equations gives:
$$2\zeta{\partial^2\eta\over \partial x^2}
+{\partial\eta\over \partial x}{\partial\zeta\over \partial x}
=3\eta^{-2/3}-3\kappa\rho_0.$$
The integrating factor of this equation is
$\partial\eta/{\partial x}$. The integration gives:
$$\zeta\bigg({\partial\eta\over \partial x}\bigg)^2
=9\eta^{1/3}-3\kappa\rho_0\eta+9\lambda~~
(\lambda~integration~constant).\eqno(18)$$
When raised to the power $3/2$, this gives:
$$\zeta^{3/2}\bigg({\partial\eta\over {\partial x}}\bigg)^3
=(9\eta^{1/3}-3\kappa\rho_0\eta+9\lambda)^{3/2}$$
If one divides (17) by this equation, $\zeta$ disappears, and it
remains the following differential equation for $\eta$:
$${2{\partial^2\eta\over \partial x^2}
\over{\big({\partial\eta\over \partial x}\big)^3}}
=-{3\kappa\gamma\eta^{1/6}
\over{(9\eta^{1/3}-3\kappa\rho_0\eta+\lambda)^{3/2}}}.$$
Here $\partial\eta/{\partial x}$ is again the
integrating factor. The integration gives:

$${2\over{\big({\partial\eta\over {\partial x}}\big)}}
=3\kappa\gamma\int{{\eta^{1/6}d\eta}
\over{(9\eta^{1/3}-3\kappa\rho_0\eta+\lambda)^{3/2}}}\eqno(19)$$
and since:
$${2\over{\delta\eta\over{\delta x}}}
={2\delta x\over{\delta\eta}}$$
through a further integration it follows:
$$x={\kappa\gamma\over {18}}\int d\eta
\int{{\eta^{1/6}d\eta}
\over{(\eta^{1/3}-{\kappa\rho_0\over
3}\eta+\lambda)^{3/2}}}.\eqno(20)$$
From here $x$ turns out as function of $\eta$, and through inversion
$\eta$ as function of $x$. Then $\zeta$ follows from (18)
and (19), and the functions $f$ through (13). Hence our
problem is reduced to quadratures.\par
\S 5. The integration constants must now be determined in such a way
that the interior of the sphere remains free from singularities
and the continuous junction to the external values of the
functions $f$ and of their derivatives at the surface of the
sphere is realised.\par
Let us put $r=r_a$, $x=x_a$, $\eta=\eta_a$, etc. at the
surface of the sphere. The continuity of $\eta$ and $\zeta$
can always be secured through a subsequent appropriate
determination of the constants $\alpha$ and $\rho$ in (14). In
order that also the derivatives stay continuous and, in keeping
with (15), $(d\eta/dx)_a=3$ and $(d\zeta/dx)_a=\eta^{-2/3}_a$,
according to (16) and (18) it must be:
$$\gamma=\rho_0\zeta^{1/2}_a\eta^{-1/6}_a,~~
\zeta_a=\eta^{1/3}_a
-{\kappa\rho_0\over 3}\eta_a+\lambda.\eqno(21)$$
From here follows:

$$\zeta_a\eta^{-1/3}_a=(f_4)_a
=1-{\kappa\rho_0\over 3}\eta_a^{2/3}+\lambda\eta^{-1/3}_a.$$
Therefore
$$\gamma=\rho_0\sqrt{(f_4)_a}.\eqno(22)$$
One sees from the comparison with (10) that in this way also the
condition $p=0$ at the surface is satisfied. The condition
$(d\eta/dx)_a=3$ yields the following determination for the
limits of integration in (19):

$${3dx\over {d\eta}}=1-{\kappa\gamma\over 6}\int^{\eta_a}_{\eta}
{{\eta^{1/6}d\eta}
\over{(\eta^{1/3}-{\kappa\rho_0\over
3}\eta+\lambda)^{3/2}}}\eqno(23)$$
and therefore (20) undergoes the following determination of
the limits of integration:

$$3(x-x_a)=\eta-\eta_a+{\kappa\gamma\over 6}
\int^{\eta_a}_{\eta}d\eta\int^{\eta_a}_{\eta}
{{\eta^{1/6}d\eta}
\over{(\eta^{1/3}-{\kappa\rho_0\over
3}\eta+\lambda)^{3/2}}}.\eqno(24)$$
The surface conditions are therefore completely satisfied. Still
undetermined are the two constants $\eta_a$ and $\lambda$, which
will be fixed through the conditions of continuity at the origin.\par
We must first of all require that for $x=0$ it should be also
$\eta=0$. If this were not the case, $f_2$ in the origin would be a
finite quantity, and an angular variation $d\phi=dx_3$ in the
origin, which in reality means no motion at all, would give a
contribution to the line element. Hence from (24) follows the
condition for fixing $\eta_a$:

$$3x_a=\eta_a-{\kappa\gamma\over 6}
\int^{\eta_a}_0d\eta\int^{\eta_a}_{\eta}
{{\eta^{1/6}d\eta}
\over{(\eta^{1/3}-{\kappa\rho_0\over
3}\eta+\lambda)^{3/2}}}.\eqno(25)$$\par
$\lambda$ will be fixed at last through the condition that the
pressure at the center of the sphere shall remain finite and
positive, from which according to (10) it follows that there $f_4$
must remain finite and different from zero. According to (13),
(18) and (23) one has:
$$f_4=\zeta\eta^{-1/3}
=\bigg(1-{\kappa\rho_0\over
3}\eta^{2/3}+\lambda\eta^{-1/3}\bigg)
\bigg[1-{\kappa\gamma\over 6}
\int^{\eta_a}_{\eta}
{{\eta^{1/6}d\eta}
\over{(\eta^{1/3}-{\kappa\rho_0\over
3}\eta+\lambda)^{3/2}}}\bigg]^2.\eqno(26)$$
One provisorily supposes either $\lambda>0$ or $\lambda<0$. Then,
for very small $\eta$:
$$f_4={\lambda\over {\eta^{1/3}}}\bigg[K+{\kappa\gamma\over 7}
{\eta^{7/6}\over{\lambda^{3/2}}}\bigg]^2,$$
where one has set:
$$K=1-{\kappa\gamma\over 6}
\int^{\eta_a}_0
{{\eta^{1/6}d\eta}
\over{(\eta^{1/3}-{\kappa\rho_0\over
3}\eta+\lambda)^{3/2}}}.\eqno(27)$$
In the center ($\eta=0$) $f_4$ will then be infinite,
unless $K=0$. But, if $K=0$, $f_4$ vanishes for $\eta=0$. In no
case, for $\eta=0$, $f_4$ results finite and different from
zero. Hence one sees that the hypothesis: either $\lambda>0$ or
$\lambda<0$, does not bring to physically practicable solutions,
and it turns out that it must be $\lambda=0$.\par
\S 6. With the condition $\lambda=0$ all the integration constants
are now fixed. At the same time the integrations to be executed
become very easy. If one introduces a new variable $\chi$ instead of
$\eta$ through the definition:
$$sin\chi=\sqrt{\kappa\rho_0\over 3}\cdot\eta^{1/3}~~~
\bigg(sin\chi_a
=\sqrt{\kappa\rho_0\over 3}\cdot\eta^{1/3}_a\bigg),\eqno(28)$$
through an elementary calculation the equations (13), (26), (10),
(24), (25) transform themselves into the following:

$$f_2={3\over{\kappa\rho_0}}sin^2\chi,~~
f_4=\bigg({{3cos\chi_a-cos\chi}\over 2}\bigg)^2,~~
f_1f^2_2f_4=1.\eqno(29)$$
$$\rho_0+p=\rho_0{2cos\chi_a\over
{3cos\chi_a-cos\chi}}\eqno(30)$$
$$3x=r^3=\bigg({\kappa\rho_0\over 3}\bigg)^{-3/2}
\bigg[{9\over 4}cos\chi_a\big(\chi-{1\over 2}sin2\chi\big)
-{1\over 2}sin^3\chi\bigg].\eqno(31)$$
The constant $\chi_a$ is determined by the density $\rho_0$ and
by the radius $r_a$ of the sphere according to the relation:

$$\bigg({\kappa\rho_0\over 3}\bigg)^{3/2}r^3_a
={9\over 4}cos\chi_a\big(\chi_a-{1\over 2}sin2\chi_a\big)
-{1\over 2}sin^3\chi_a.\eqno(32)$$
The constants $\alpha$ and $\rho$ of the solution for the external
region come from (14):

$$\rho=\eta_a-3x_a~~~\alpha=\eta^{1/3}_a-\zeta_a$$
and obtain the values:

$$\rho=\bigg({\kappa\rho_0\over 3}\bigg)^{-3/2}
\bigg[{3\over 2}sin^3\chi_a-
{9\over 4}cos\chi_a\big(\chi_a-{1\over 2}sin2\chi_a\big)
\bigg]\eqno(33)$$
$$\alpha=\bigg({\kappa\rho_0\over 3}\bigg)^{-1/2}
\cdot sin^3\chi_a.\eqno(34)$$
{\it When one avails of the variables $\chi$, $\vartheta$, $\phi$
instead of $x_1$, $x_2$, $x_3$ ($ix$), the line element in the
interior of the sphere takes the simple form:}
$$ds^2=\bigg({{3cos\chi_a-cos\chi}\over 2}\bigg)^2 dt^2
-{3\over{\kappa\rho_0}}[d\chi^2+sin^2\chi d\vartheta^2
+sin^2\chi sin^2\vartheta d\phi^2].\eqno(35)$$
{\it Outside the sphere the form of the line element remains the
same as in ``Mass point'':}
$$\eqalign{ds^2&=(1-\alpha/R)dt^2-{dR^2\over{1-\alpha/R}}
-R^2(d\vartheta^2+\sin^2\vartheta
d\phi^2)\cr
&where~~R^3=r^3+\rho.}\eqno(36)$$
Now $\rho$ will be determined by (33), while for the
mass point it was $\rho=\alpha^3$.\par
\S 7. The following remarks apply to the complete solution of
our problem contained in the previous paragraphs .\par
1. The spatial line element ($dt=0$) in the interior of the sphere
reads:

$$-ds^2={3\over{\kappa\rho_0}}[d\chi^2+sin^2\chi d\vartheta^2
+sin^2\chi sin^2\vartheta d\phi^2].$$\par
This is the known line element of the so called non Euclidean
geometry of the spherical space. {\it Therefore the geometry of the
spherical space holds in the interior of our sphere}. The
curvature radius of the spherical space will be
$\sqrt{3/\kappa\rho_0}$. Our sphere does not constitute the whole
spherical space, but only a part, since $\chi$ can not grow up to
$\pi/2$, but only up to the limit $\chi_a$. For the Sun the
curvature radius of the spherical space, that rules the geometry
in its interior, is about 500 times the radius of the Sun (see
formulae (39) and (42)).\par
That the geometry of the spherical space, that up to now had to be
considered as a mere possibility, requires to be real in the
interior of gravitating spheres, is an interesting result of
Einstein's theory.\par
Inside the sphere the quantities:

$$\sqrt{3\over {\kappa\rho_0}}d\chi,~~
\sqrt{3\over {\kappa\rho_0}}sin\chi d\vartheta,~~
\sqrt{3\over {\kappa\rho_0}}sin\chi sin\vartheta d\phi,\eqno(37)$$
are ``naturally measured'' lengths. The radius ``measured inside''
from the center of the sphere up to its surface is:

$$P_i=\sqrt{3\over {\kappa\rho_0}}\chi_a.\eqno(38)$$
The circumference of the sphere, measured along a meridian (or
another great circle) and divided by $2\pi$, is called the
radius ``measured outside'' $P_o$. It turns out to be:

$$P_o=\sqrt{3\over {\kappa\rho_0}} sin \chi_a.\eqno(39)$$
According to the expression (36) of the line element outside the
sphere this $P_o$ is clearly identical with the value
$R_a=(r^3_a+\rho)^{1/3}$ that the variable $R$ assumes at the
surface of the sphere.\par
With the radius $P_o$ one gets for $\alpha$ from (34) the simple
relations:

$${\alpha\over {P_o}}=sin^2\chi_a,~~
\alpha={\kappa\rho_0\over 3}P^3_o.\eqno(40)$$
The volume of our sphere is:

$$\eqalign{V&=\bigg(\sqrt{3\over {\kappa\rho_0}}\bigg)^3
\int^{\chi_a}_0 d\chi sin^2 \chi\int^{\pi}_0 d\vartheta\sin\vartheta
\int^{2\pi}_0 d\phi\cr
&=2\pi\bigg(\sqrt{3\over {\kappa\rho_0}}\bigg)^3
\bigg(\chi_a-{1\over 2}sin 2\chi_a\bigg).}$$
Hence the mass of our sphere will be ($\kappa=8\pi k^2$)

$$M=\rho_0V={3\over {4k^2}}\sqrt{3\over {\kappa\rho_0}}
\bigg(\chi_a-{1\over 2}sin 2\chi_a\bigg).\eqno(41)$$\par
2. About the equations of motion of a point of infinitely small
mass outside our sphere, which maintain tha same form as in ``Mass
point'' (there equations (15)-(17)), one makes the following
remarks:\par
For large distances the motion of the point occurs according to
Newton's law, with $\alpha/2k^2$ playing the r\^ole of the
attracting mass. Therefore $\alpha/2k^2$ can be designated as
``gravitational mass'' of our sphere.\par
If one lets a point fall from the rest at infinity down to the surface
of the sphere, the ``naturally measured'' fall velocity takes the
value:

$$v_a={1\over\sqrt{1-\alpha/R}}{dR\over ds}
=\sqrt{\alpha\over R_a}.$$
Hence, due to (40):

$$v_a=sin\chi_a.\eqno(42)$$\par
For the Sun the fall velocity is about 1/500 the velocity of light.
One easily satisfies himself that, with the small value thus resulting for
$\chi_a$ and $\chi$ ($<\chi_a$), all our equations coincide with
the equations of Newton's theory apart from the known second order
Einstein's effects.\par
3. For the ratio between the gravitational mass $\alpha/2 k^2$ and
the substantial mass $M$ one finds

$${\alpha\over {2k^2M}}={2\over 3}
{sin^3\chi_a\over{\chi_a-{1\over 2}sin 2\chi_a}}.\eqno(43)$$\par
With the growth of the fall velocity $v_a$ ($=sin\chi_a$), the
growth of the mass concentration lowers the ratio between the
gravitational mass and the substantial mass. This becomes clear
for the fact that {\it e. g.} with constant mass and increasing
density one has the transition to a smaller radius with emission
of energy (lowering of the temperature through radiation).\par
4. The velocity of light in our sphere is

$$v={2\over {3cos\chi_a-cos\chi}},\eqno(44)$$
hence it grows from the value $1/cos\chi_a$ at the surface to the value
$2/(3cos\chi_a-1)$ at the center. The value of the pressure
quantity $\rho_0+p$ according to (10) and (30) grows in direct
proportion to the velocity of light.\par
At the center of the sphere ($\chi=0$) velocity of light
and pressure become infinite when $cos\chi_a=1/3$, and the
fall velocity becomes $\sqrt{8/9}$ of the (naturally measured)
velocity of light. Hence there is a limit to the
concentration, above which a sphere of incompressible fluid can
not exist. If one would apply our equations to values
$cos\chi_a<1/3$, one would get discontinuities already outside the
center of the sphere. One can however find solutions of the
problem for larger $\chi_a$, which are continuous at least outside
the center of the sphere, if one goes over to the case of
either $\lambda>0$ or $\lambda<0$, and satisfies the condition
$K=0$ (Eq. 27). On the road of these solutions, that are clearly
not physically meaningful, since they give infinite pressure at
the center, one can go over to the limit case of a mass
concentrated to one point, and retrieves then the relation
$\rho=\alpha^3$, which, according to the previous study, holds for
the mass point. It is further noticed here that one can speak of a
mass point only as far as one avails of the variable $r$, that
otherwise in a surprising way plays no r\^ole for the geometry and
for the motion inside our gravitational field. {\it For an
observer measuring from outside it follows from (40) that a sphere of
given gravitational mass $\alpha/2k^2$ can not have a radius
measured from outside smaller than:}

$$P_o=\alpha.$$
{\it For a sphere of incompressible fluid the limit will be
$9/8\alpha$.} (For the Sun $\alpha$ is equal to 3 km, for a mass
of 1 gram is equal to $1.5\cdot10^{-28}$ cm.)
\end